\documentclass[aps,prl,showpacs,twocolumn,superscriptaddress]{revtex4}
\usepackage{bm,color}
\usepackage{graphicx}
\usepackage{amsmath, amssymb}

\begin{document}
\title{Machine-learning detection of the Berezinskii--Kosterlitz--Thouless transitions in the $q$-state clock models}
\author{Yusuke Miyajima}
\author{Yusuke Murata}
\author{Yasuhiro Tanaka}
\author{Masahito Mochizuki}
\affiliation{Department of Applied Physics, Waseda University, Okubo, Shinjuku-ku, Tokyo 169-8555}
\begin{abstract}
We demonstrate that a machine learning technique with a simple feedforward neural network can sensitively detect two successive phase transitions associated with the Berezinskii--Kosterlitz--Thouless (BKT) phase in $q$-state clock models simultaneously by analyzing the weight matrix components connecting the hidden and output layers. We find that the method requires only a data set of the raw  spatial spin configurations for the learning procedure. This data set is generated by Monte-Carlo thermalizations at selected temperatures. Neither prior knowledge of, for example, the transition temperatures, number of phases, and order parameters nor processed data sets of, for example, the vortex configurations, histograms of spin orientations, and correlation functions produced from the original spin-configuration data are needed, in contrast with most of previously proposed machine learning methods based on supervised learning. Our neural network evaluates the transition temperatures as $T_2/J=0.921$ and $T_1/J=0.410$ for the paramagnetic-to-BKT transition and BKT-to-ferromagnetic transition in the eight-state clock model on a square lattice. Both critical temperatures agree well with those evaluated in the previous numerical studies.
\end{abstract}
\maketitle

\section{INTRODUCTION}
Machine learning techniques have been widely used for image recognition and are applied practically nowadays~\cite{LeCun15,Goodfellow16}. The techniques are based on a design of devices capable of giving correct guesses, assignments, and classifications of unknown data by extracting features and patterns that the data contain. This procedure resembles research in the field of physics, in which physical principles are often deduced from accumulated experimental data. Therefore, machine learning techniques might have good compatibility with scientific research and have thus been applied to a wide variety of physical issues~\cite{Carleo17,Carleo19,Ohtsuki20,Nomura17}. An ultimate goal of this research direction is to discover new physics through machine learning, but state-of-the-art machine-learning-based research in physics has not yet reached this stage. Currently, we are developing efficiencies of the techniques by testing whether the techniques correctly guess and identify well-known behaviors and properties of established theoretical models. 

In particular, several spin models have been intensively examined to test whether the machine learning techniques can detect and classify phases and their phase transitions. To date, the Ising model~\cite{Tanaka17,Carrasquilla17,Arai18,Suchsland18,Hu17,Wetzel17,Wang16,Ponte17,Nieuwenburg17,Liu18} and the XY model~\cite{Richter-Laskowska18,Beach18,ZhangW19,Shiina20,Rodriguez-Nieva19} in two dimensions have been the subject of intensive studies. The former model exhibits the second-order phase transition to a symmetry-broken phase with decreasing temperature; that is, the paramagnetic-to-(anti)ferromagnetic transition where the order parameter is well defined. The latter model possesses a continuous U(1) symmetry and thus does not exhibit a normal second-order phase transition to a symmetry-broken phase according to the Mermin--Wagner theorem~\cite{MerminWagner}. Instead, the XY model exhibits the Berezinskii--Kosterlitz--Thouless (BKT) transition or the vortex-binding transition associated with topological defects; that is, a transition from a high-temperature phase having unbound vortices and antivortices to a low-temperature phase having bound vortex--antivortex pairs with decreasing temperature~\cite{Kosterlitz73,Kosterlitz74,Tobochnik79}.

Machine learning techniques are mainly classified into two categories: techniques based on supervised learning and techniques based on unsupervised learning. The supervised learning techniques use training data, to which labels of desired output are attached, whereas the unsupervised learning techniques treat the cases without labels and the desired outputs are unknown. To date, not only supervised learning techniques~\cite{Tanaka17,Carrasquilla17,Arai18,Suchsland18} but also many different unsupervised learning techniques, such as principle-component analyses combined with the autoencoder method~\cite{Hu17,Wetzel17} or clustering analysis~\cite{Wang16}, the use of support-vector machines~\cite{Ponte17}, learning by confusion~\cite{Nieuwenburg17}, and the use of discriminative cooperative networks~\cite{Liu18}, have succeeded in detecting the second-order phase transition in the Ising model. 

In contrast, only a few machine learning techniques have succeeded in detecting the BKT transition in the XY model~\cite{Richter-Laskowska18,Beach18,ZhangW19,Shiina20,Rodriguez-Nieva19}. Most of these techniques are based on supervised learning techniques~\cite{Richter-Laskowska18,Beach18,ZhangW19,Shiina20} and require feature engineering in advance; that is, preprocessed data of, for example, vortex configurations~\cite{Beach18}, histograms of spin orientations~\cite{ZhangW19}, and spin correlation functions~\cite{Shiina20} need to be prepared as input data instead of the raw  data of spatial spin configurations. Moreover, most of these techniques need prior knowledge of fundamental properties of the model, such as approximate values of transition temperatures~\cite{Richter-Laskowska18} and the number of phases~\cite{Richter-Laskowska18,ZhangW19}. These aspects are problematic for the establishment of a generalized scheme for the machine-learning-based study of phase-transition phenomena.

In this paper, to establish a machine learning technique for detecting general phases and phase transitions including topological phase transitions, we extend and generalize a method proposed in Refs.~\cite{Tanaka17,Arai18}. Those works demonstrated that the second-order phase transition of the Ising model and that of the transverse-field Ising model can be detected in the weight matrix components of a convolution neural network. We apply this technique to the eight-state clock model and examine if the technique is capable of detecting the BKT phase and the BKT transitions in this model. We use unprocessed spin-configuration data generated through Monte-Carlo thermalization at various temperatures as training data and use temperatures at which the Monte-Carlo thermalization is performed as labels of the training data for the supervised learning. By analyzing the weight matrix after training using a newly introduced correlation function, we demonstrate that this method with a simple feedforward neural network can detect three different phases including the BKT phase and two successive BKT transitions in the eight-state clock model simultaneously. The proposed method is found to have several advantages over previously proposed machine-learning-based methods as listed below.
\begin{itemize}
\item The proposed method is applicable not only to phase transitions to a symmetry-broken phase but also to topological phase transitions with no trivial order parameters.
\item The proposed method can detect multiple phases and multiple phase transitions simultaneously.
\item The proposed method can detect phase transitions without any prior knowledge of the model, such as knowledge of the transition temperatures, the number of phases, or order parameters.
\item The proposed method does not require processed data created through feature engineering and requires only the raw  data of spatial spin configurations as input.
\end{itemize}

\section{MODEL}
\begin{figure} 
\includegraphics[scale=1.0]{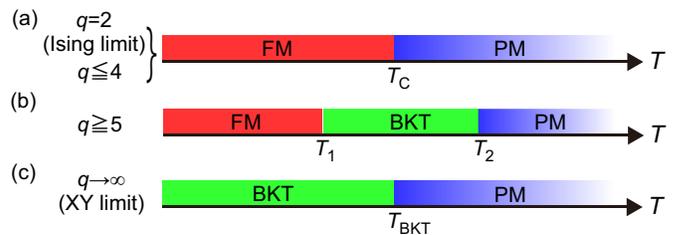}
\caption{Phase diagrams of $q$-state clock models as a function of temperature $T$. (a) When $q\le4$, the model exhibits a single second-order phase transition from the paramagnetic (PM) phase to the ferromagnetic (FM) phase. (b) When $q\ge5$, the model exhibits two phase transitions from the PM phase to the BKT phase to the FM phase. (c) In the continuum limit of $q$$\rightarrow$$\infty$, the model exhibits a single BKT transition from the PM phase to the BKT phase.}
\label{Fig01}
\end{figure}
We study ferromagnetic ($J>0$) $q$-state clock models on a square lattice. The Hamiltonian is given by
\begin{eqnarray}
\mathcal{H}=-J \sum_{\langle i,j \rangle} \bm{S}_i \cdot \bm{S}_j.
\end{eqnarray}
The planar discrete classical spin vectors are
\begin{eqnarray}
\bm{S}_i=\left(S_i^x, S_i^y\right)
=\left(\cos \frac{2 \pi k}{q}, \sin \frac{2 \pi k}{q} \right),
\end{eqnarray}
where $k=0, 1, \cdots , q-1$. Here, the summation is taken over pairs of adjacent sites $i$ and $j$. These models have been recognized as important because they connect two fundamental classical spin models; that is, the Ising model ($q$=2) in the discrete limit and the XY model ($q$$\rightarrow$$\infty$) in the continuum limit (see Fig.~\ref{Fig01}). The Ising model exhibits the second-order phase transition at a finite temperature even in two dimensions because of the discretized spin variables. In contrast, the XY model exhibits the BKT transition, which is regarded as a topological phase transition, at a finite temperature. The critical integer number $q_{\rm c}$ above which the BKT transition emerges has been a subject of controversy~\cite{Jose78,Elitzur79,Tobochnik82,Lapilli96,Hwang09,Baek10,Kumano13}. A recent study has put an end to this controversy~\cite{Kumano13}, which concluded that a single second-order transition from the paramagnetic phase to the ferromagnetic phase occurs when $q\leq 4$, whereas two phase transitions from the paramagnetic phase to the BKT phase to the ferromagnetic phase occur when $q\geq 5$ as claimed in the earliest stage of the research~\cite{Jose78,Elitzur79}.

\begin{figure} 
\includegraphics[scale=1.0]{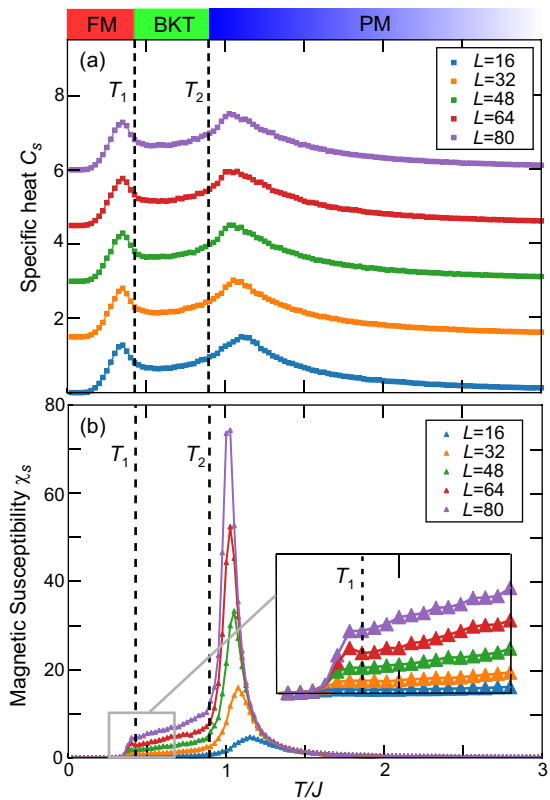}
\caption{Temperature profiles of (a) specific heat $C$ and (b) magnetic susceptibility $\chi_s$ for the eight-state clock model on a square lattice calculated using the Monte-Carlo technique for various system sizes of $L \times L$. Two critical temperatures $T_1$ and $T_2$ for successive phase transitions among the paramagnetic (PM), BKT, and ferromagnetic (FM) phases with decreasing temperature are indicated by dashed lines. Note that anomalies (cusps, peaks, and jumps) in these physical quantities do not correspond to the BKT phase transition points.}
\label{Fig02}
\end{figure}
The specific heat cannot be used for the determination of the critical temperatures of the BKT transitions, although this quantity is usually exploited to detect phase transitions. Figure~\ref{Fig02}(a) shows calculated temperature profiles of the specific heat of the $q$-state clock model with $q$=8 for various system sizes of $L\times L$; these profiles were obtained through Monte-Carlo calculations. The profiles have two broad peaks. However, these peaks do not represent the critical temperatures ($T_1$ and $T_2$)~\cite{Tobochnik82,Lapilli96,Hwang09}. Here, $T_1$ is the critical temperature for the phase transition between the low-temperature ferromagnetic phase and the intermediate BKT phase, whereas $T_2$ is that for the phase transition between the BKT phase and the high-temperature paramagnetic phase. The calculated temperature profiles of magnetic susceptibility are presented in Fig.~\ref{Fig02}(b). A sharp peak at higher temperature and a weak jump at lower temperature are observed. The determination of the critical temperatures from these anomalies is, in principle, possible but difficult. It requires highly accurate evaluations of the anomaly points and precise size-scaling analysis because significant finite-size effects due to the logarithmic corrections hamper the conventional finite-size scaling scheme such as the Binder plot.

\begin{figure*} 
\includegraphics[scale=1.0]{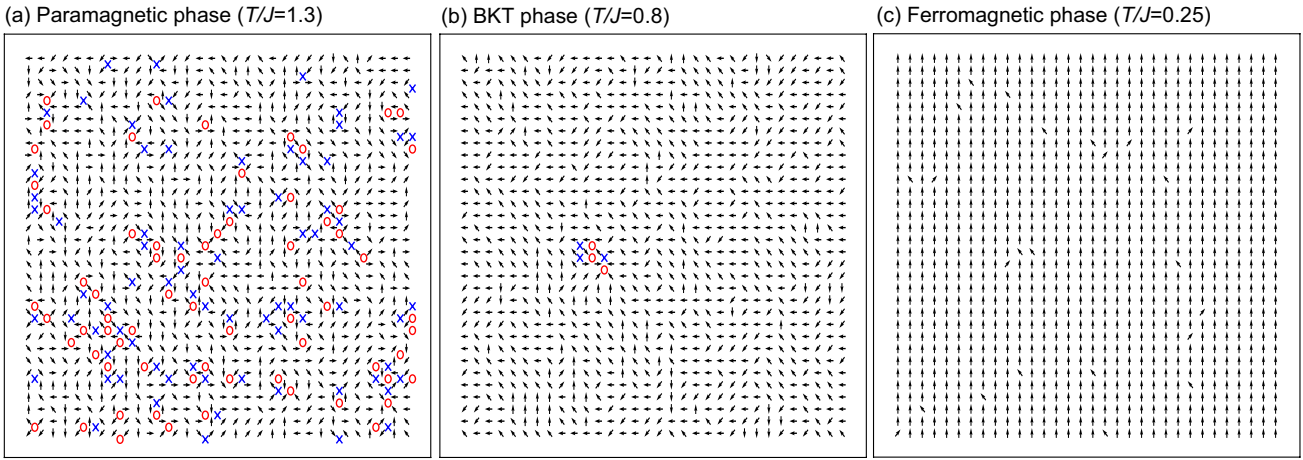}
\caption{(a--c) Typical snapshots of spatial spin configurations of (a) the paramagnetic phase at $T/J$=1.3, (b) the BKT phase at $T/J$=0.8, and (c) the ferromagnetic phase at $T/J$=0.25 for the eight-state clock model on a square lattice of $32 \times 32$ sites generated by the Monte-Carlo thermalization. The positions of vortices and antivortices are respectively indicated by (red) circles and (blue) crosses. There are many unbound vortices and antivortices in the paramagnetic phase [(a)], whereas there are three vortex--antivortex bound pairs in the BKT phase [(b)]. In the ferromagnetic phase [(c)], the spins are aligned in a nearly parallel manner.}
\label{Fig03}
\end{figure*}
Figure~\ref{Fig03} presents typical snapshots of spatial spin configurations of (a) the paramagnetic phase at $T/J$=1.3, (b) the BKT phase at $T/J$=0.8, and (c) the ferromagnetic phase at $T/J$=0.25 for the eight-state clock model obtained through Monte-Carlo thermalization for a system of $32 \times 32$ sites. The positions of vortices and antivortices are respectively indicated by (red) circles and (blue) crosses. Notably, there are many unbounded vortices and antivortices in the paramagnetic phase (a) whereas there are a few vortex--antivortex bounded pairs in the BKT phase (b). In the ferromagnetic phase (c), a nearly uniform spin configuration is observed, where almost all the spins are aligned in a parallel manner. It is difficult to distinguish these phases and their phase transitions with human eyes. We demonstrate that a simple fully connected neural network can correctly distinguish the physical phases and precisely determine the phase-transition points. 

\section{METHOD AND RESULTS}
We first prepare many spin configurations of the $q$-state clock model for several differently sized lattices of $L \times L$ sites ($L$ = 16, 32, 48, 64, 72, and 80) using the single-flip Monte-Carlo technique based on the Metropolis algorithm. We perform the Monte-Carlo thermalization with 200,000 iterative steps at various temperatures of $T_n=T_0+(n-1)\Delta T$ with $T_0$=0.01, $\Delta T$=0.01, and $n=1, \cdots, 300$; i.e., at 300 temperature points between $T_{n=1}$=0.01 and $T_{n=300}$=3.00 with the same interval of $\Delta T$. We prepare 10 thermalized spin configurations for each temperature point.

\begin{figure} 
\includegraphics[scale=1.0]{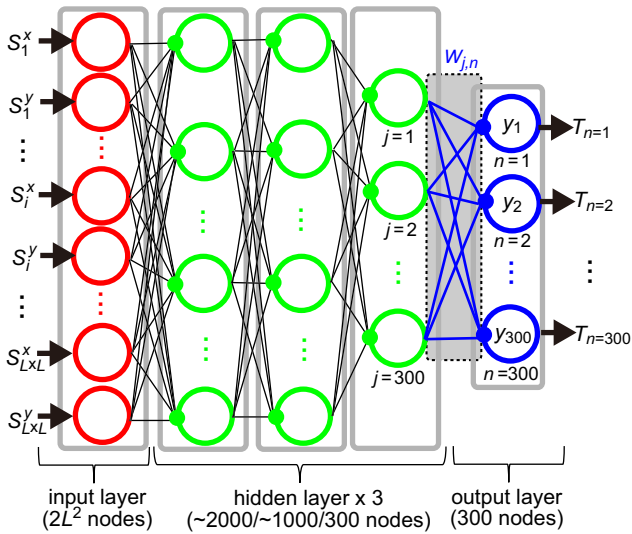}
\caption{Schematic diagram of the fully connected neural network used in the simulations for a system size of $L \times L$.}
\label{Fig04}
\end{figure}
We implement a fully connected neural network with the machine learning library KERAS~\cite{KERAS15} using TENSORFLOW~\cite{TENSF15a,TENSF15b} as the computational backend. Our neural network comprises one input layer, three hidden layers, and one output layer of neurons [see Fig.~\ref{Fig04}]. The two adjacent layers are fully connected, where each neuron in one layer is connected to every neuron in the previous and next layers. The interlayer connections are represented by sets of correlation weights called weight matrices. The inputs to the neural network are the snapshot spin configurations obtained through the Monte-Carlo thermalization; that is, the $x$ and $y$ components of the planar discretized spin vectors $\bm S_i=(S_i^x, S_i^y)$ at all sites ($i=1, 2, \cdots, L^2$) on the square lattice. Therefore, the required number of nodes in the input layer is $2L^2$. Meanwhile, the answer labels are given by the so-called one-hot representation of temperatures. The temperature $T_n=T_0+(n-1)\Delta T$ ($n=1, \cdots, 300$) is represented by a vector $\bm t$ with 300 components in which only the $n$th component is set to unity while all other components are set to zero. In other words, when the temperature is $T=T_n$, the $k$th component $t_k$ ($k=1,2, \cdots, 300$) of the vector $\bm t$ is given by $t_k=\delta_{k,n}$, where $\delta_{i,j}$ is Kronecker's delta. Each node in the output layer is assigned to one of the components of the vector $\bm t$, and the number of nodes in the output layer is thus 300. Typical numbers of nodes of the three hidden layers are 2000, 1000, and 300. As the activation function, the rectified linear unit (ReLU) is used for the first and third hidden layers whereas the softmax function is used for the second hidden layer and the output layer.

Using the prepared $300 \times 10$ sets of spatial spin configurations as training data, we optimize the weights of our neural network so as to correctly guess temperatures at which the input spin configuration is obtained. More concretely, the weight matrices are optimized through backpropagation to minimize the cost function on the training data~\cite{LeCun15}. Here, the cross-entropy error function is employed as the cost function:
\begin{equation}
E(\bm t^i, \bm y^i)=\sum_i \sum_k t_k^i \log_e y_k^i,
\end{equation}
where $i$ is the index of the spin-configuration data sets ($1\le i \le 3000$), $k$ is the index of the output nodes or the discretized temperature points ($1\le k \le 300$), and $y_k^i$ is the output value at the $k$th output node for the $i$th training. Here, $t_k^i$ is the $k$th component of the vector $\bm t^i$ in the one-hot representation of the temperature at which  the spin configuration data used for the $i$th training are generated through Monte-Carlo thermalization. A training algorithm called Adam is used to train the neural network through minimizing the cost function~\cite{Kingma14}.

After completing the training, we focus on the weights $W_{j,n}$ connecting the $j$th node in the last (the third in the present study) hidden layer ($j=1,2, \cdots, 300$) and the $n$th node in the output layer ($n=1,2, \cdots, 300$)~\cite{Tanaka17,Arai18}. It is found that the weight matrix $W_{j,n}$ has distinct behaviors among different physical phases, which enables us to distinguish the phases and to detect transitions among the phases. 
\begin{figure} 
\includegraphics[scale=0.5]{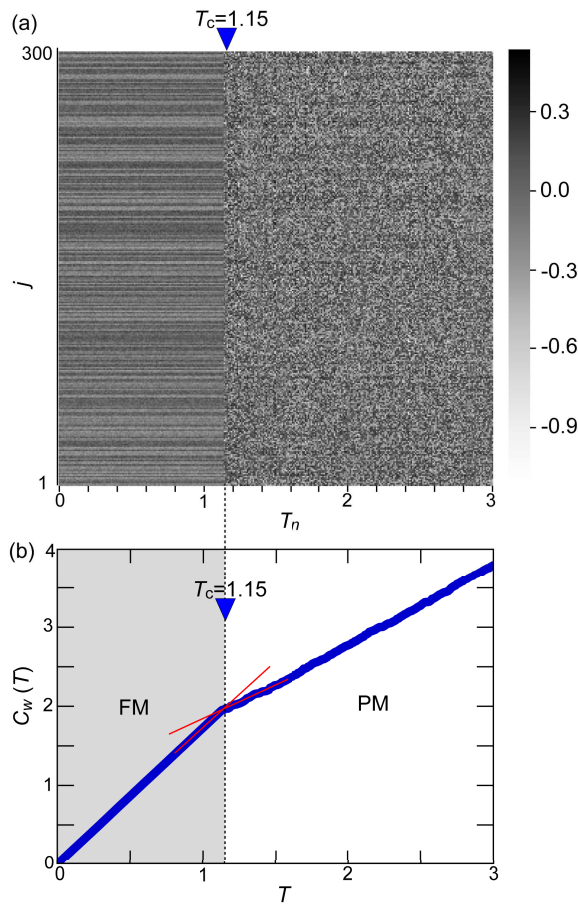}
\caption{(a) Heat map of the weight matrix components $W_{j,n}$ connecting the $j$th node in the last hidden layer and the $n$th node in the output layer in the plane of $j(=1,2, \cdots, 300$ and $T_n=T_0+(n-1)\Delta T$ ($n=1, 2, \cdots, 300$) for the four-state clock model ($q$=4) on the square lattice of $L=80$, where $T_0$=0.01 and $\Delta T$=0.01. A clear change in pattern at $T_{\rm c}=1.15$ is seen and ascribed to the phase transition from the paramagnetic (PM) phase to the ferromagnetic (FM) phase at this temperature. (b) Correlation $C_W(T)$ of the weight matrix components $W_{j,n}$ defined by Eq.~(\ref{eq:Cw}) as a function of $T_n$, clearly showing a change in slope at $T_{\rm c}$.}
\label{Fig05}
\end{figure}
\begin{figure} 
\includegraphics[scale=0.5]{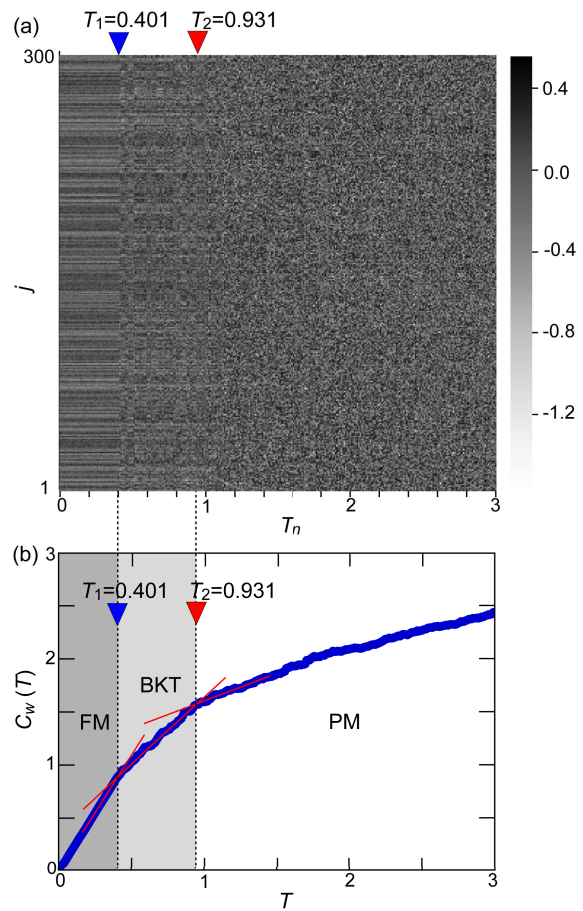}
\caption{(a) Heat map of the weight matrix components $W_{j,n}$ for the eight-state clock model ($q$=8) on the square lattice of $L=72$. (b) Correlation $C_W(T)$ defined by Eq.~(\ref{eq:Cw}) as a function of $T_n$, clearly showing two slope changes attributable to two successive phase transitions from the paramagnetic (PM) phase to the BKT phase to the ferromagnetic (FM) phase.}
\label{Fig06}
\end{figure}
To confirm the efficiency of the proposed method, we first examine the four-state clock model ($q$=4) on the square lattice, which exhibits a single second-order phase transition from the paramagnetic phase to the ferromagnetic phase with decreasing temperature. Figure~\ref{Fig05}(a) presents a grayscale mapping of amplitudes of the weight matrix $W_{j,n}$ for a system size of $L$=80. Here, the horizontal axis represents the label $n$ of nodes in the output layer or the corresponding temperatures $T_n$ while the vertical axis represents the label $j$ of nodes in the last hidden layer. There is a clear difference in pattern between the areas above and below $T_{\rm c}=1.15$. More specifically, we see a horizontal-stripe pattern appears in the low-temperature regime and a sandstorm pattern in the high-temperature regime. These two regimes are assigned to different phases (i.e., the ferromagnetic phase and paramagnetic phase, respectively) in the present model. The evaluated boundary $T_{\rm c}$ of 1.15 coincides well with the exact transition temperature of $T_{\rm c}=J/\ln(1+\sqrt{2})\approx1.1346J$ in the thermodynamic limit~\cite{Onsager44,Brush67}. 

To analyze the pattern changes in $W_{j,n}$ more quantitatively, we introduce the quantity
\begin{equation}
C_W(T)=\frac{1}{N_{\rm h}}\sum_{m (T_m<T)}\sum_{j=1}^{N_{\rm h}}W_{j,m}W_{j,m+1},
\label{eq:Cw}
\end{equation}
where $N_{\rm h}$(=300) is the number of nodes on the last hidden layer. The product $W_{j,m}W_{j,m+1}$ quantifies the extent of similarity/difference between the adjacent columns in the weight matrix $W_{j,n}$, and the quantity $C_W(T)$ is a sum of the products over $m$ for $T_m<T$. This quantity can reveal pattern changes associated with phase transitions. Figure~\ref{Fig05}(b) presents the temperature profile of $C_W(T)$, which changes slope at the transition points $T_{\rm c}$. Although the pattern change in the heat map of Fig.~\ref{Fig05}(a) is easy to identify in the present case, there might be cases in which the pattern change is difficult to identify by eyes. In fact, the newly introduced quantity $C_W(T)$ can be used to detect the pattern changes in the heat map sensitively as will be argued for the next example.

We next examine the eight-state clock model ($q$=8). This model exhibits successive two phase transitions associated with the BKT phase; that is, phase transitions from the paramagnetic phase to the BKT phase to the ferromagnetic phase with decreasing temperature. Figure~\ref{Fig06}(a) shows the heat map of $W_{j,n}$ for a system size of $L$=72. We expect two pattern changes in the heat map corresponding to the two BKT phase transitions. We see a clear pattern change at $T_1=0.421$ associated with the BKT-to-ferromagnetic phase transition at lower temperature. In this heat map, however, another pattern change associated with the paramagnetic-to-BKT phase transition is difficult to identify with human eyes. Thus, we calculate the correlation function $C_W(T)$ for this heat map. Figure~\ref{Fig06}(b) shows the temperature profile of $C_W(T)$, in which the solid (red) lines are presented for guides to eyes. We find that three phases characterized by distinct slopes appear. The two variations of slope reflect successive two BKT phase transitions. We can determine the second critical temperature $T_2$ for the paramagnetic-to-BKT phase transition in this profile.

\begin{table}[h]
\caption{Critical temperatures of the eight-state clock model for various system sizes. The case of $L\rightarrow \infty$ corresponds to the thermodynamic limit.}
\begin{tabular}{c||cccccc|c} 
\hline
$L$ & 16 & 32 & 48 & 64 & 72 & 80 & $\infty$ \\
\hline 
$T_1$ &  0.319 & 0.376 & 0.388 & 0.392 & 0.401 & 0.390 & 0.410 \\
$T_2$ &  1.279 & 1.000 & 0.943 & 0.931 & 0.931 & 0.930 & 0.921 \\
\hline
\end{tabular}
\label{tab:T1T2}
\end{table}

\begin{figure} 
\includegraphics[scale=1.0]{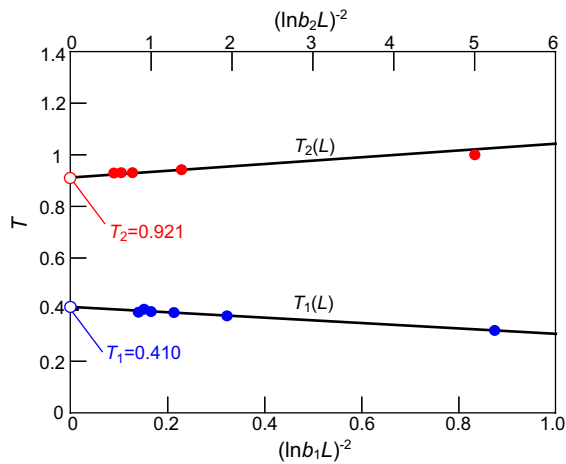}
\caption{System-size scaling of the critical temperatures $T_1$ and $T_2$ for the eight-state clock model. Extrapolations indicate $T_1$=0.410 and $T_2$=0.921 in the thermodynamic limit (see the main text).}
\label{Fig07}
\end{figure}
The critical temperatures $T_1$ and $T_2$ depend on the system size (see Table~\ref{tab:T1T2}). We perform a finite-size scaling analysis to evaluate the critical temperatures in the thermodynamic limit. According to the Kosterlitz--Thouless theory, the correlation length $\xi$ is proportional to $\exp(c/\sqrt{t})$, where $c$ is a constant and $t$ is the relative temperature, which is given by
\begin{eqnarray}
t=\frac{|T-T_{\rm BKT}(\infty)|}{T_{\rm BKT}(\infty)}.
\end{eqnarray}
Here, $T_{\rm BKT}(\infty)$ is the transition temperature in the thermodynamic limit. Because $\xi$ is maximized at $T=T_{\rm BKT}(L)$ to be $\xi=L$ in finite-sized systems, the size dependence of $T_{\rm BKT}(L)$ is given by
\begin{eqnarray}
T_{\rm BKT}(L) = T_{\rm BKT}(\infty) \pm \frac{c^2 T_{\rm BKT}(\infty)}{(\ln bL)^2},
\label{eq:FSS}
\end{eqnarray}
where the $+$ ($-$) sign is taken for the higher (lower) critical temperature $T_2$ ($T_1$).

\begin{table}[h]
\caption{Critical temperatures of the eight-state clock model in the thermodynamic limit obtained by the present and previous studies.}
\begin{tabular}{ccc} \hline
Reference & $T_1$ & $T_2$ \\ \hline 
Ref.~\cite{Tomita02} & 0.4259(4) & 0.8936(7) \\
Ref.~\cite{Brito10}  & 0.42(2)   & 0.898(7)  \\
Present Work & 0.410 & 0.921 \\ \hline
\end{tabular}
\label{tab:TKT}
\end{table}
The size dependencies of the critical temperatures $T_1$ and $T_2$ are plotted in Fig.~\ref{Fig07}. According to the finite-size scaling analysis with Eq.~(\ref{eq:FSS}), $T_1$ and $T_2$ in the thermodynamic limit are evaluated to be $T_1 = 0.410$ and $T_2 = 0.921$. These values are consistent with those evaluated in previous studies (see Table~\ref{tab:TKT})~\cite{Tomita02,Brito10}. Here, the values of $b$ and $c$ obtained by fitting are ($b$, $c$) = (0.182, $-0.104$) for $T_1$ and ($b$, $c$) = (0.049, 0.016) for $T_2$.

\section{Discussion}
In this section, we discuss the proposed method and the analysis scheme in detail.
\subsection{Comparison with Monte-Carlo methods}
The present machine-learning-based method has both advantages and disadvantages over the Monte-Carlo methods for the computational studies on the phase-transition phenomena. The advantage is, in fact, the reduced computing cost. The present method requires neither a large number of Monte-Carlo samplings to obtain thermal averages of physical quantities nor a time-consuming thermalization procedure to realize thermal equilibrium states in contrast to the Monte-Carlo methods. Instead, the present method requires a set of spin configurations as training data, which are prepared by the Monte-Carlo thermalization procedure. Importantly, this procedure does not require long time because it does not need to reach real thermal equilibrium. Indeed, we stop the thermalization cycles at a small iteration number of $2 \times 10^5$ irrespective of the system size. Even a training with not fully relaxed spin configurations turned out to be efficient to detect the BKT transitions possibly because the spin configurations taken on the way to the thermal equilibrium already involve features of the equilibrium state at which the system would finally arrive. It is also important to mention that the neural network requires a training procedure, but it turned out to take only little computational time (a few ten minutes or less even for the largest system size of $L=80$), which is much shorter than the time-scale of the standard Monte-Carlo calculations (typically several hours or even a few days). 

On the contrary to the above advantage, the proposed machine learning method has some disadvantages over the Monte-Carlo methods at the same time, e.g., difficulty to obtain quantitatively accurate results for critical phenomena. This might be because the present method is based on a kind of image-recognition technique, and hence the controlled improvement of the accuracy is difficult. This problem is not specific to the present method but is common to most of the machine learning methods in contrast to the Monte-Carlo calculations in which a larger number of Monte-Carlo samplings can necessarily suppress the statistical errors of thermal averages. Note that in the case of machine learnings, on the contrary, an excess training often causes an overlearning problem. Another disadvantage is difficulty to distinguish anomaly points associated with real phase transitions in the plot of $C_W(T)$. Indeed, we found additional kinks in some plots at higher temperatures which are not related with real phase transitions. We consider that the present method based on the image recognition senses not only phase transitions but also cross-over phenomena. To distinguish real phase transitions from several anomalous behaviors of the system, a more elaborate analysis may be required.

As argued above, the present machine-learning method has both advantages and disadvantages over the Monte-Carlo methods. However, we should note that the purpose of this work is not to propose a method alternative to or better than the Monte-Carlo methods but to demonstrate the possibility of neural networks in detection of topological phases and related topological transitions, which are characterized neither by trivial order parameters nor by simple symmetry breakings. One of the goals of the research aiming at application of machine learnings to the physical science is discovering a novel physics. A state-of-the-art research, however, has not necessarily reached this level yet. Currently, we are working to accumulate experiences and benchmarks of the machine-learning techniques and to improve their methodologies through examining established mathematical models and well-known physical situations. At this stage, it is of essential importance to clarify what the machine learnings can do and/or cannot do. Indeed, several types of machine-learning detection of the BKT transitions in the clock models have been reported in literatures. The present work has achieved a precious contribution to development of the machine-learning research because the capability of detecting the BKT phase and the BKT transitions have been confidently demonstrated with novel advantages and improvement over the previously proposed machine-learning-based methods as argued in the end of the introduction section.

\subsection{Neural network and weight matrix}
In the present work, the weight matrix $W_{ij}$ connecting the last hidden layer and the output layer is a 300$\times$300 matrix irrespective of the system size. The output of the present neural network is not a phase type but a temperature to which the input spin configuration belongs, in contrast to most of the previously proposed machine-learning-based methods. The number of output nodes (i.e., 300) is determined by the number of temperature meshes or how finely to divide the temperature range into meshes, which directly affects precision of the evaluation of transition temperatures. Because of this, the present method requires a rather large number of nodes for the output layer. But, in return, the method can be widely implemented without any prior knowledge about models, e.g., the number of phases and the expecting phase types. 

On the other hand, we employ 300 nodes in the last hidden layer right before the output layer irrespective of the system size. In these nodes, features of the input spin configuration are accumulated in an integrated manner after processed by the neural network. Because we treat systems of up to $L=80$, which has 80$\times$80$\times$2=12,800 input data, we need, at least, 300 nodes to achieve a proper functionality of the neural network. We think that a neural network with a small number of hidden nodes fails to capture the features of spin configuration because of excessive coarse graining. This 300$\times$300 weight matrix is not huge at all as compared with those used for, e.g., image recognitions, voice recognitions, and character recognitions, and it does not need much computational cost. Indeed, we trained the neural network to a satisfactory level only within a few ten or less minutes even for the largest system of $L=80$ in the present work.

\subsection{Data analysis}
For evaluations of the transition temperatures, we first visually guess the kink positions in the plot of $C_W(T)$. This human intervention does not cause uncertainty in the evaluation of transition temperatures so much. We perform the linear fittings of finite regions in front and back of the guessed kink position. We subsequently confirm that the two fitting lines are indeed crossed on the kink position. In this procedure, the evaluation of slope is not affected by roughness of the plot of $C_W(T)$ because we calculate an averaged slope in a finite region before and behind the kink area. With this procedure, we can eventually suppress errors due to ambiguity of the visually guessed kink position. Indeed, we did several trials with differently trained neural networks for the same system size. The plot of $C_W(T)$ and the extent of change in slope vary for every trial. But we confirmed that the evaluated kink positions fluctuate only negligibly, which guarantees that the transition temperatures are correctly evaluated with very little ambiguity.

It should be mentioned that we did not perform any uncertainty estimations in the present work. In fact, a method of the uncertainty estimation for neural networks has not been established yet. A fundamental difficulty lies in the fact that conventional techniques of the uncertainty estimation for continuous stochastic processes cannot be applied to neural networks with high-dimension data input. The modeling and estimate of the uncertainty of the neural networks used in the present work are an issue for future challenges.

\section{CONCLUSION}
We demonstrated that a simple fully connected neural network is capable of detecting the two BKT phase transitions in the $q$-state clock model on a square lattice in the weight matrix $W_{j,n}$ connecting the hidden layer and the output layer after training of the supervised learning procedure. The newly introduced correlation function $C_W(T)$ is powerful in the detection of BKT phase transitions through the identification of the pattern change in the greyscale map (i.e., the heat map) of the weight matrix components $W_{j,n}$ in the plane of $j$ (i.e., the index of nodes in the last hidden layer) and $n$ (i.e., the index of nodes in the output layer or the temperature points). We evaluated the critical temperatures $T_1$ and $T_2$ of the two BKT transitions in the eight-state clock model by conducting finite-size-scaling analyses and found that their values in the thermodynamic limit are consistent with those reported in the literature.

\begin{figure} 
\includegraphics[scale=1.0]{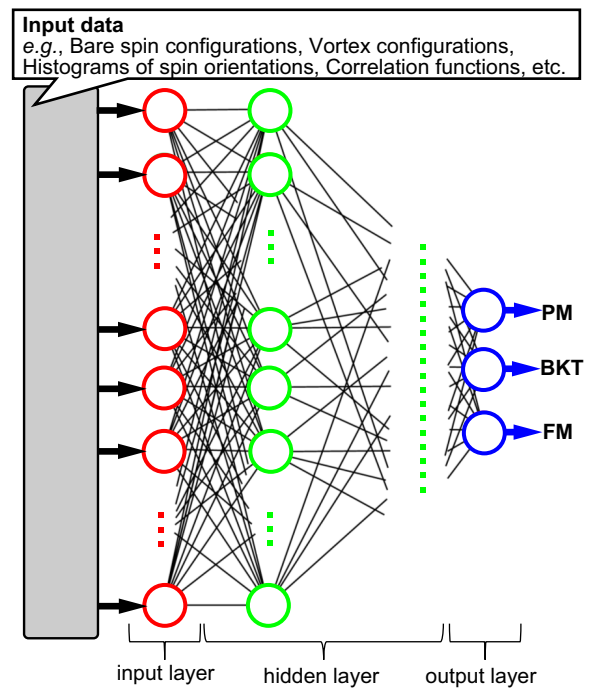}
\caption{Typical neural network of the supervised machine learning for spin-phase classifications where the input data are the raw  spin configurations or preprocessed data of, for example, vortex configurations, spin-orientation histograms, or correlation functions whereas the output data are the names of possible phases, such as the paramagnetic (PM) phase, BKT phase, and ferromagnetic (FM) phase.}
\label{Fig08}
\end{figure}
The present method has several advantages over previous methods. A typical neural network of the supervised machine learning for spin-phase classification has the structure shown in Fig.~\ref{Fig08}. Here, the input data are the bare spin configurations or the preprocessed data of, for example, vortex configurations, spin-orientation histograms, or correlation functions created through feature engineering whereas the output data are the names of possible phases, such as the paramagnetic phase, BKT phase, and ferromagnetic phase. To train this neural network, we need to know the name of the phase to which each input datum belongs as an answer label. The approximate values of critical temperatures and the number of phases must therefore be known in advance. Meanwhile, the present method does not require preprocessed data, and only data sets of the raw  spin configurations, which can be easily generated in Monte-Carlo simulations, are used as input data. Moreover, the answer labels attached to the input data are only the temperatures at which the spin configurations are generated. These temperatures are absolutely known without uncertainty because they are set manually  in the Monte-Carlo simulations. Thus, prior knowledge of critical temperatures is not needed. Furthermore, because we determine the number of phases and the phase-transition points through heat-map analysis, prior knowledge of the number of phases is not required either.

Finally, we would like to discuss an issue of future challenge. As mentioned in the introduction section, the clock models are a class of fundamental classical spin models which connect two limiting cases, i.e., the Ising limit ($q$=2) and the $XY$ limit ($q\rightarrow\infty$). In the present study, we chose the eight-state clock model as a target of the machine learning analysis, which exhibits rather well-defined three spin phases, i.e., the paramagnetic, BKT, and ferromagnetic phases. On the contrary, the five-state clock model, which is considered to be located in-between these two limits, must be difficult to treat. It had been long debated whether there exists the BKT phase in this model. The detection of the narrow BKT phase in this model is necessarily difficult but worth trying. We may need to improve both the neural network and the training data with e.g., more nodes in the hidden layer, more thermalization steps in the training-data preparation, and more sets of high-quality training data to sharpen the kinks and anomalies in the correlation function $C_W(T)$ at the transition points. 

\section{Appendix}
\begin{figure*} 
\includegraphics[scale=0.5]{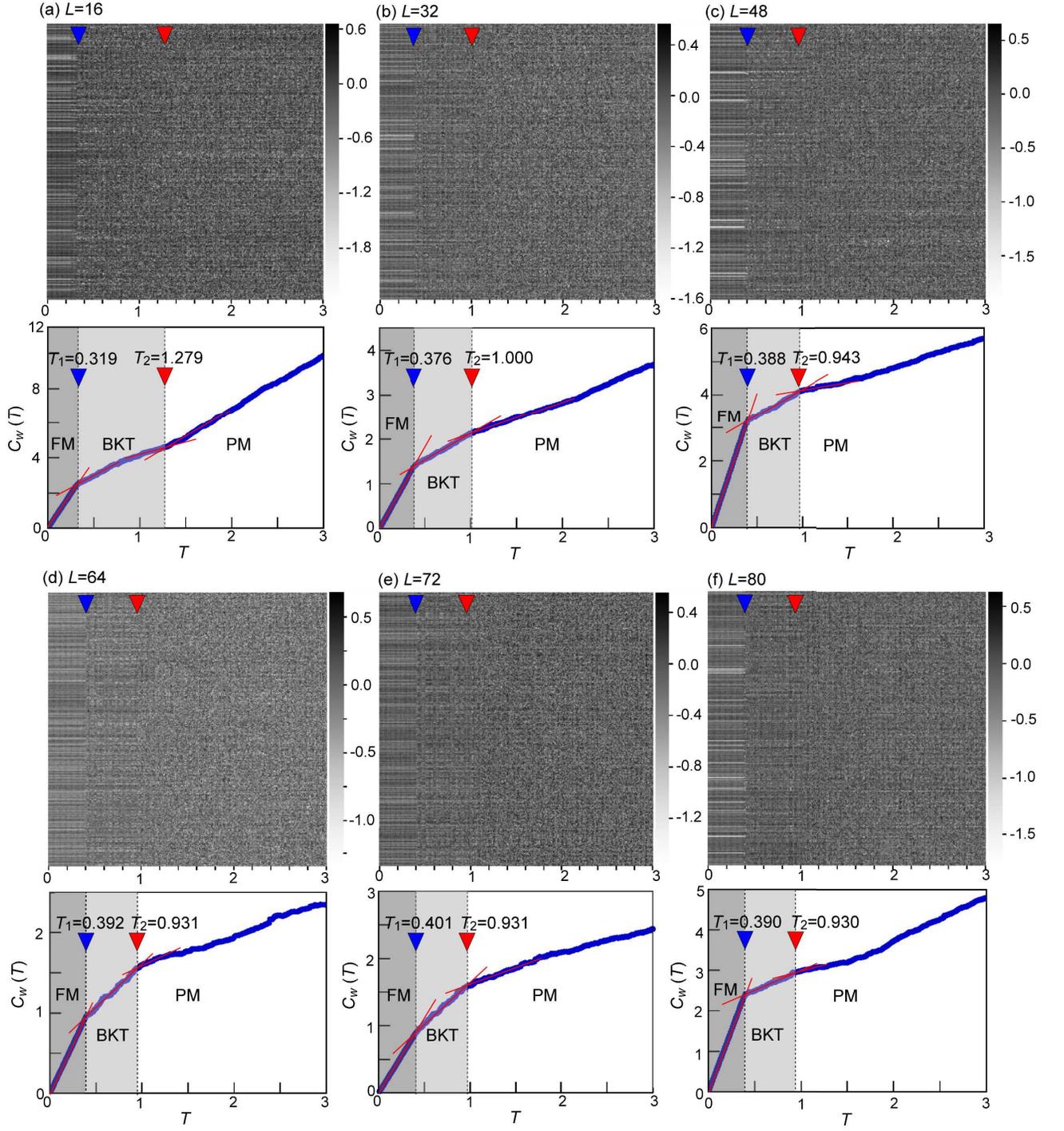}
\caption{Heat maps and corresponding temperature profiles of $C_W(T)$ of the eight-state clock model for various lattice sizes. Solid (red) lines in the plots of $C_W(T)$ are visual guides that distinguish the three phases: the paramagnetic (PM) phase, BKT phase, and ferromagnetic (FM) phase. The crossing points are considered phase transition points.}
\label{Fig09}
\end{figure*}
Figure~\ref{Fig09}(a)--(d) presents heat maps obtained using our neural network after training and the temperature profiles of the corresponding correlation function $C_W(T)$ for various lattice sizes. The solid (red) lines in the plots of $C_W(T)$ are visual guides that distinguish the three physical phases (i.e., the paramagnetic phase, BKT phase, and ferromagnetic phase) according to the variations of slopes in the linear fitting. The crossing points are considered transition points. Note that when the lattice size is as small as $L=16$, the temperature profile of $C_W(T)$ cannot be fitted using a single linear line in the intermediate area, whereas the profile in the paramagnetic area at higher temperature and that in the ferromagnetic area at lower temperature can be fitted with linear functions. There seems to be an additional point at which the slope of the linear fitting changes in the intermediate area. This anomaly of $C_W(T)$ might be attributed to an artifact generated by the smallness of the lattice size because it disappears for the larger lattices. In the present analysis, the transition temperatures $T_1$ and $T_2$ for $L=16$ are assigned to the points at which the slopes of the linear fittings in the higher- and lower-temperature areas change.

The weight matrix after training is never unique but varies for every trial upon the variations of training data, training procedure, and several other conditions of the neural network. The slope and amplitude of the correlation function $C_W(T)$ also vary. Nevertheless, we find a tendency in the temperature profile of $C_W(T)$ in Fig.~\ref{Fig09} that the changes of slope are always negative both at $T_1$ and $T_2$ for all the system sizes except for the change at $T_2$ for $L$=16 i.e., the smallest system size [Fig.~\ref{Fig09}(a)]. We should also note that an additional kink appears at a higher temperature in some plots [see e.g., Figs.~\ref{Fig09}(b) and (f)], although no phase transition takes place at the corresponding temperature in reality. A further analysis is needed to clarify what the neural network indeed detects and captures, which is an issue left for future study.

In the meanwhile, the plots of $C_W(T)$ exhibit slight fluctuation and roughness, but they are never a troublesome for our analysis. The present method is not a scheme based on a statistical averaging so that the quantity at each temperature point necessarily fluctuates. Despite the fluctuations, the characteristics of each phase should appear over the temperature range in which the phase emerges. Thus, we focus on the slope of $C_W(T)$ evaluated by the linear fitting, which is expected to capture the averaged and global features of each phase. Importantly, the slight fluctuation and roughness in the plot of $C_W(T)$ have only negligible influence on the results of linear fittings.

We also realize that the kink at $T_2$ is relatively unclear in some plots of $C_W(T)$, particularly, for the largest system of $L=80$ [Fig.~\ref{Fig09}(f)]. In fact, the clearness or sharpness of the kink can be improved rather easily. This issue is related with the numbers of thermalization steps and the nodes in the hidden layer. In the present work, we adopted the same number of thermalization steps (i.e., $2 \times 10^5$) and the same number of nodes (i.e., 300) in the last hidden layer for all the system sizes from $L=16$ to $L=80$ to equalize the conditions as much as possible. In fact, these choices do not have any peculiar reasons. We simply selected, more or less, equal and unbiased conditions, aiming at the equal-footing demonstrations, while a neural network has a numerous number of parameters to be tuned or given by hand. We did neither examine systematic variations of the neural-network parameters and the Monte-Carlo parameters nor discuss results obtained for different parameters because there are too many variable parameters for this machine-learning-based research, and we are afraid that their case studies would obscure a purpose of the work. However, we have examined a lot of cases with different numbers of thermalization steps and hidden nodes. Through these examinations, we have found that a larger number of thermalization steps tend to give a clearer and sharper change in slope at $T_2$. This indicates that a larger number of thermalization steps are required for preparing good training data when the system size is larger, as is also the case for the Monte-Carlo calculations. We have also found that a neural network with a larger number of hidden nodes tends to give a clearer kink at $T_2$. In the present neural network, the $2L^2$ input data are integrated into 300 nodes in the last hidden layer after processed through the hidden layers. The above finding indicates that too much integration or suppression of data might cause a failure in capturing the features of each physical phase, and an appropriate number of hidden nodes should be employed depending on the system size. 

\section{Acknowledgments}
\noindent
MM thanks M. Imada, T. Ohtsuki, R. Pohle, Y. Nomura, and Y. Yamaji for valuable discussions and comments. MM is supported by Japan Society for the Promotion of Science KAKENHI (Grant No. 16H06345, No. 19H00864, No. 19K21858 and No. 20H00337), CREST, the Japan Science and Technology Agency (Grant No. JPMJCR20T1), a Research Grant in the Natural Sciences from the Mitsubishi Foundation, and a Waseda University Grant for Special Research Projects (Project No. 2020C-269). YT is supported by Japan Society for the Promotion of Science KAKENHI (Grant No. 19K23427 and No. 20K03841). 

\end{document}